\documentclass[11pt]{article}
\usepackage{amsmath,amssymb,bm}
\usepackage{amsthm}
\usepackage{algorithm}
\usepackage{algpseudocode}
\usepackage{graphicx}
\usepackage{subcaption}
\usepackage[margin=1.25in,marginparwidth=1.1in,marginparsep=3pt]{geometry}
\usepackage{hyperref}

\usepackage{xcolor}
\usepackage{booktabs}

\newcommand{\E}{\mathbb{E}}
\newcommand{\Sx}{\Sigma_{\mathrm{x}}}
\newcommand{\Sy}{\Sigma_{\mathrm{y}}}
\newcommand{\Stx}{\Sigma_{\mathrm{\tilde{x}}}}
\newcommand{\Sty}{\Sigma_{\mathrm{\tilde{y}}}}

\newcommand{\mx}{\mu_{\mathrm{x}}}
\newcommand{\my}{\mu_{\mathrm{y}}}
\newcommand{\tx}{\tilde{\mathrm{x}}}
\newcommand{\ty}{\tilde{\mathrm{y}}}

\newcommand{\R}{\mathbb{R}}

\newcommand{\mcX}{\mathcal{X}}
\newcommand{\mcN}{\mathcal{N}}
\DeclareMathOperator{\diag}{diag}

\DeclareMathOperator{\var}{var}

\DeclareMathOperator{\cov}{cov}

\newcommand{\mbfu}{\mathbf{u}}
\newcommand{\mbfz}{\mathbf{z}}

\newcommand{\htheta}{\hat{\theta}}

\newcommand{\ptri}{p_{\mathrm{tri}}}
\newcommand{\pduo}{p_{\mathrm{duo}}}
\newcommand{\ktri}{\hat k^{\mathrm{tri}}}

\newcommand{\notkappa}{\text{\tiny$\sim$}\kappa}
\newcommand{\notk}{\text{\tiny$\sim$}k}

\newtheorem{claim}{Claim}
\newtheorem{lemma}{Lemma}
\newtheorem{corollary}{Corollary}
\newtheorem*{definition}{Definition}
\newtheorem{remark}{Remark}

\usepackage{authblk}

\title{Fast likelihood evaluation for multivariate sensory discrimination models}
\author[1]{Philip Greengard}
\author[2]{Fangfang Hong}
\author[2]{David H. Brainard}
\author[3,1]{Alex Williams}
\affil[1]{Center for Computational Neuroscience, Flatiron Institute}
\affil[2]{Department of Psychology, University of Pennsylvania}
\affil[3]{Center for Neural Science, New York University}
\date{\today}

\begin{document}
\maketitle

\begin{abstract}
Sensory discrimination is often measured by fitting statistical models with likelihood functions that are computationally challenging to evaluate. In the standard triangle and duo-trio tests, a subject is presented with three stimuli, two of one kind and one of another, and is asked to identify the odd one out (the triangle test) or which of two stimuli matches a reference (the duo-trio test). In a Thurstonian model, the subject's internal representation of each stimulus is a Gaussian whose mean and covariance depend on the stimulus, and the probability that the subject's response is correct is an integral over a $3d$-dimensional Gaussian with no closed form for $d > 1$, where $d$ is the dimension of the stimulus space. To fit a statistical model to data collected from triangle or duo-trio tests, the likelihood function, and often its gradient, need to be evaluated many times, making statistical inference inefficient. Here, we introduce fast algorithms for evaluating certain classes of likelihood functions for the triangle and duo-trio tests. The algorithms rely on reducing the dimension of these functions. The triangle probability, a function of $2d + d(d+1)$ arguments, reduces analytically to a function of $2d$ arguments, and the duo-trio probability, when the two stimuli have equal covariances, reduces to a function of a single scalar for any $d$. After reducing the dimensionality of those functions, we use neural networks and Chebyshev expansions to evaluate the lower-dimensional functions efficiently. Our algorithms demonstrate orders of magnitude speedups over existing approaches at comparable accuracy, both for a single evaluation and when used to compute the maximum a posteriori (MAP) estimate of a model of color discrimination. The dimensionality reduction we use generalizes to non-Gaussian noise models with elliptical level sets, and to tests with more than three stimuli.
\end{abstract}

\newpage
\section{Introduction}\label{sec:intro}
Sensory discrimination experiments have a long history in, for example, color
perception~\cite{macadam1942}, taste~\cite{fisher1935}, and the perception of
weight~\cite{fechner1860}. 
In sensory discrimination experiments, the triangle test and the duo-trio test are standard choices
for measuring perceptual differences between two stimuli~\cite{mullen1988}.
In the triangle test, a subject is given three stimuli, two of which are the same and the
third different. The subject is asked to identify the stimulus that differs from the other two.
In the duo-trio test, a subject is first presented with a reference stimulus. The
subject is then presented with two further stimuli, one is the same as the reference
and the other different. The subject is asked to identify which of the two
stimuli is the same as the reference.
These tests appear in a variety of industrial and academic research settings. For example, 
the color discrimination task of \cite{hong2026} is a triangle test. In public health research, 
similar sensory discrimination tasks are run for identifying perceptual differences in 
taste~\cite{saavedra2015, mcmahon2016, kim2010, hartley2022} and olfaction~\cite{hummel1997, hummel2007}.
In industrial settings the triangle and duo-trio tests are frequently used for quality control
\cite{iso10399, iso4120}.

\begin{figure}[t]
\centering
\includegraphics[width=0.8\textwidth]{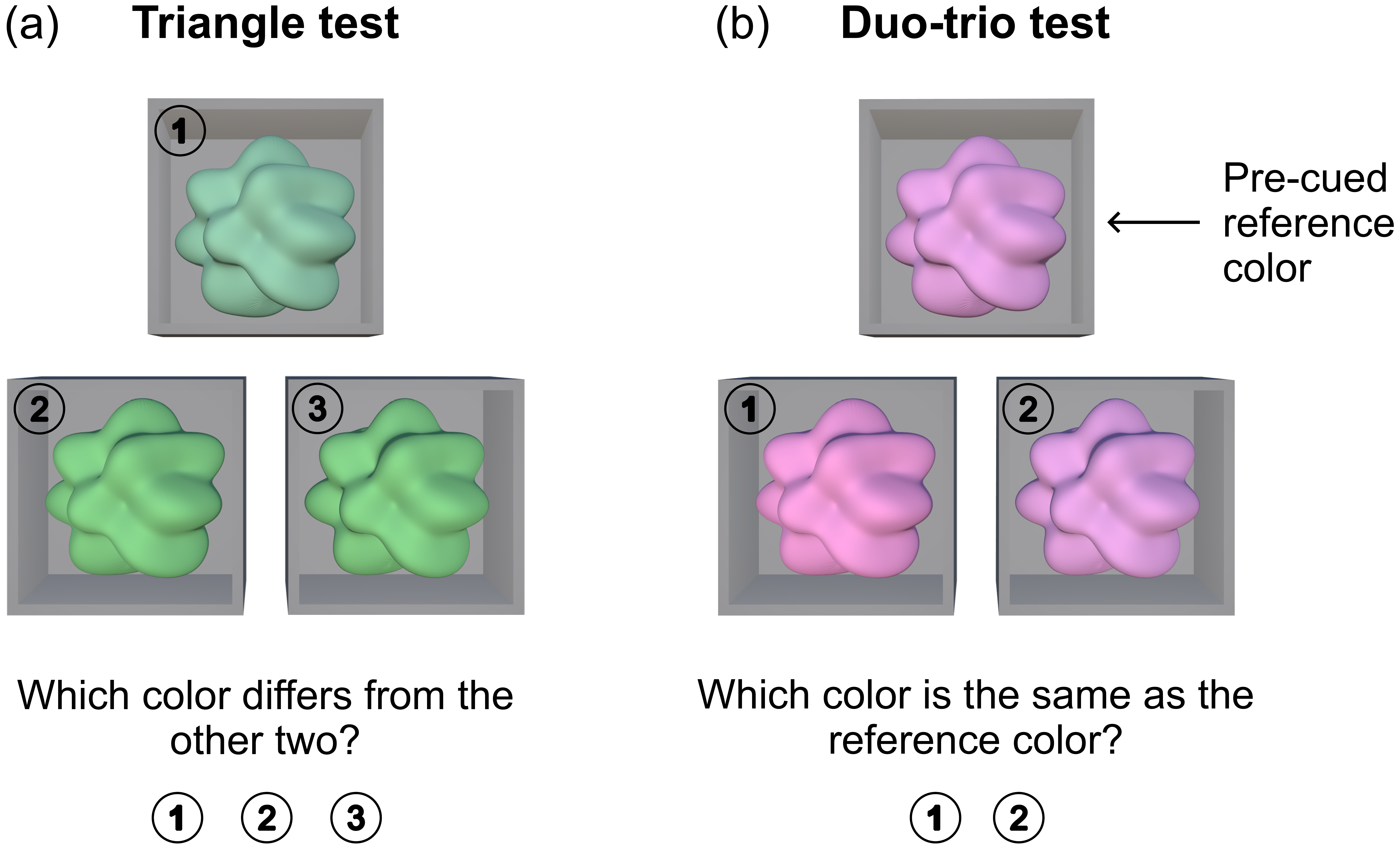}
\caption{Examples of color discrimination experiments with the (a) triangle test from 
\cite{hong2026} and (b) duo-trio test from \cite{morimoto2026}.}
\label{fig:trials}
\end{figure}

In general, the space of stimuli is $\mcX \subseteq \R^d$ for some $d$, such as
in color tasks where $d \leq 3$~\cite{hong2026}. Researchers use tests like the triangle and duo-trio 
tests, in combination with statistical models, to estimate how sensory perceptual discrimination varies 
as a function of stimulus space. 
The most common approach, dating to Thurstone~\cite{thurstone1927} and underlying signal detection theory~\cite{green1966,macmillan2005detection}, models each stimulus presentation as producing a noisy internal representation drawn from a Gaussian distribution.
The multivariate generalization of this model is well-established~\cite{ennis1993, mullen1987, ashby1986varieties}, and has been applied across a variety of domains~\cite{amazeen1999perceptual,fitousi2013variants,soto2015general,hong2026}.
However, calculating choice probabilities under multivariate Thurstonian models of triad tasks is nontrivial and computationally intensive, which has likely limited their use.

\begin{figure}[t]
\centering
\includegraphics[width=\textwidth]{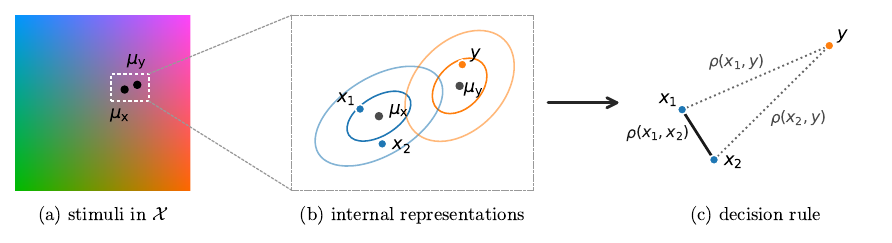}
\caption{Schematic of the Thurstonian model and decision rule in the triangle test. 
(a) the repeated stimulus $\mx$ and the unique stimulus $\my$ in the stimulus space $\mcX$. (b) the subject's internal representations of stimuli, $x_1, x_2 \sim \mcN(\mx, \Sx)$ and
$y \sim \mcN(\my, \Sy)$. Ellipses are level sets of the Gaussian probability density function. 
(c) the decision rule in which the subject chooses the
draw not in the closest pair of points under $\rho$, here $y$.}
\label{fig:thurstonian}
\end{figure}

In the triangle test, where two of the three presented stimuli are the same, 
we denote by $x_1, x_2 \sim \mcN(\mx, \Sx)$, the subject's independent internal
representations of the repeated stimulus. We denote by $y \sim \mcN(\my, \Sy)$, independent of $x_1, x_2$, 
the internal representation of the unique stimulus. The subject's response to the triangle 
test, their guess of which 
is the unique stimulus, is modeled as the stimulus whose 
internal representation is farthest from the other two, where distance is measured by some   
distance function $\rho: \R^d \times \R^d \to \R^+$. That is, the probability of a correct response in the 
triangle test, which we denote by $\ptri$, is
\begin{equation}\label{eq:p_triangle}
\ptri(\mx, \my, \Sx, \Sy) = \Pr \bigl(
      \rho(x_1, x_2) < \min\{\rho(x_1, y),\, \rho(x_2, y)\}
    \bigr).
\end{equation}
Figure \ref{fig:thurstonian} illustrates the model for one trial.

In the duo-trio test, the subject indicates which of the two presented stimuli is closer to the 
reference stimulus. That is, we define  
$x_{\mathrm{ref}} \sim \mcN(\mx, \Sx)$ to be the subject's internal representation of the reference stimulus,
$x_1 \sim \mcN(\mx, \Sx)$ a repeated copy of the reference stimulus and $y_1 \sim \mcN(\my, \Sy)$ the 
other stimulus, all independent. Then $\pduo$, the probability of a correct response in the duo-trio
test, is given by
\begin{align}\label{eq:p_duotrio}
\pduo(\mx, \my, \Sx, \Sy) = \Pr \big( \rho(x_{\mathrm{ref}}, x_1) < \rho(x_{\mathrm{ref}}, y_1) \big).
\end{align}

An important consideration in the model is the choice of distance function
$\rho$ of \eqref{eq:p_triangle}, \eqref{eq:p_duotrio}. The models of \cite{mullen1987,
kapenga1987, mullen1988} choose $\rho$ to be Euclidean distance, which amounts to 
assuming that the stimulus space is parameterized such that perceptual distance is Euclidean. 
For $d \geq 2$, such a parameterization exists only in instances in which 
the metric $\Sigma(x)^{-1}$ has zero Riemann curvature \cite{docarmo1992}, where 
$\Sigma(x)$ is the covariance of the internal representation of stimulus $x$.
Even in cases where the curvature does vanish, the parameterization that makes $\rho$
Euclidean is determined by the covariance field $\Sigma(x)$, which is the quantity being 
estimated, and thus is unavailable to the experimenter in advance. Indeed, in the 
coordinates in which stimuli are presented, discrimination often varies with the reference 
stimulus \cite{macadam1942, hong2026}, indicating that Euclidean $\rho$ is not the 
appropriate distance function in that representation.

The color discrimination models of \cite{hong2026, morimoto2026} use $\rho$ of the form
\begin{align}\label{eq:dist}
  \rho(x, y)
  = (x - y)^\top S^{-1} (x - y),
\end{align}
where $S$ is a convex combination of $\Sx$ and $\Sy$. 
This is a natural choice of $\rho$ because (i) the stimuli $\mx, \my$ presented to subjects are 
close enough that $\Sx, \Sy$ are approximately equal, and (ii) for any $\Sx, \Sy$, 
the ideal observer \cite{green1966,geisler2011} (in the sense described in Appendix \ref{sec:ml_rule}) compares internal
representations $u, v$ through $(u - v)^\top \Sx^{-1} (u - v)$.

In many experiments with triangle and duo-trio tests, one of the main computational challenges 
in analyzing experimental data is the accurate and efficient evaluation of the Bernoulli 
probabilities \eqref{eq:p_triangle} 
and \eqref{eq:p_duotrio}. These functions are integrals of the same integrand over different regions
 of $\R^{3d}$. Specifically,
 \begin{align}
  \ptri(\mx, \my, \Sx, \Sy)
    &= \int_{S_{\mathrm{tri}}}
       \phi(u_1; \mx, \Sx)\, \phi(u_2; \mx, \Sx)\, \phi(v; \my, \Sy)
       \,\mathrm{d}u_1\, \mathrm{d}u_2\, \mathrm{d}v,
  \label{eq:f_tri_int} \\
  \pduo(\mx, \my, \Sx, \Sy)
    &= \int_{S_{\mathrm{duo}}}
       \phi(u_{1}; \mx, \Sx)\, \phi(u_2; \mx, \Sx)\, \phi(v; \my, \Sy)
       \,\mathrm{d}u_{1}\, \mathrm{d}u_2\, \mathrm{d}v ,
  \label{eq:f_duo_int}
\end{align}
where $\phi(u; \mu, \Sigma)$ denotes the Gaussian probability density function (p.d.f.) with mean 
$\mu$ and covariance $\Sigma$ and where 
\begin{align}
  S_{\mathrm{tri}} &= \bigl\{ (u_1, u_2, v) \in \R^{3d} \;:\;
      \rho(u_1, u_2) < \min\{ \rho(u_1, v),\, \rho(u_2, v) \} \bigr\},
  \label{eq:S_tri} \\
  S_{\mathrm{duo}} &= \bigl\{ (u_1, u_2, v) \in \R^{3d} \;:\;
      \rho(u_1, u_2) < \rho(u_1, v) \bigr\}.
  \label{eq:S_duo}
\end{align}
There is a large body of work on algorithms for accurately and 
efficiently evaluating functions \eqref{eq:p_triangle}, \eqref{eq:p_duotrio} 
in the case where $\rho$ is Euclidean distance
\cite{ennis1993, brockhoff2010, sensR}. 
In one dimension ($d=1$), when the variance is assumed constant for all $x \in \mcX$, the functions
\eqref{eq:p_triangle}, \eqref{eq:p_duotrio} can be efficiently evaluated --- \eqref{eq:p_triangle} is a one-dimensional 
integral with smooth integrand and \eqref{eq:p_duotrio} is a second-order polynomial in error 
functions~\cite{brockhoff2010, sensR}. 
For $d > 1$, evaluation of the probabilities $\ptri, \pduo$ has no closed form and often requires
computationally costly numerical methods each time the likelihood function is evaluated. Quadrature methods 
are introduced for $\ptri$ in \cite{ennis1986, kapenga1987} and for both $\ptri, \pduo$ in \cite{mullen1988}.
For the duo-trio probability, \cite{ennis1994} uses several mathematical and numerical observations
to reduce the integral to a series that can be evaluated accurately and efficiently. 
Monte Carlo is an alternative approach for both $\ptri$ and $\pduo$ \cite{ennis1985, ennis1986, hong2026, morimoto2026} and has a standard error of $\mathcal{O}(n^{-1/2})$ with $n$ samples.

Here, we introduce fast algorithms for evaluating $\ptri,\pduo$ for $d > 1$ where $\rho$ is of the form 
\eqref{eq:dist}.
The main mathematical observation
underlying these algorithms is that $\ptri$ and $\pduo$ are invariant under 
affine transformations of the stimulus space. That is, for any invertible $A$ and any
$b \in \R^d$, we have
\begin{align}\label{eq:invariant_trans}
\ptri(\mx, \my, \Sx, \Sy) &= \ptri(A\mx + b, A\my + b, A \Sx A^\top, A \Sy A^\top), \\
\pduo(\mx, \my, \Sx, \Sy) &= \pduo(A\mx + b, A\my + b, A \Sx A^\top, A \Sy A^\top).
\end{align}
As a result, $\ptri$ and $\pduo$, which are functions of $2d + d(d+1)$ arguments, can be 
reduced to functions of $2d$ variables (see Claim \ref{claim:linear_invariance}).
In particular, $d^2$ variables are removed by 
an appropriate choice of invertible $A$ and another $d$ by translation of the means.\footnote{We note that this reduction of the dimensionality of $\ptri, \pduo$ is not possible with the assumption
of Euclidean $\rho$ for any covariances $\Sx, \Sy$ as in \cite{ennis1994}.
For Euclidean $\rho$, the functions $\ptri, \pduo$ are invariant under the 
transformations \eqref{eq:invariant_trans} where $A$ is a scalar multiple of an orthogonal matrix. 
Using an appropriate $A$, we can reduce $\ptri, \pduo$ by $d(d - 1) / 2 + 1$ arguments, the dimensionality 
of scalar multiples of orthogonal $d \times d$ matrices. After translation, and transformation by $A$, 
we have a function of $2d + d(d + 1)/2 - 1$ arguments.}

After this transformation, we approximate the transformed $\ptri$, a function of $2d$ 
arguments, with a neural network with $369$ parameters for $d = 2$ and $401$ parameters 
for $d = 3$.
We train each network on $3 \times 10^5$ points labeled by Monte Carlo with $n = 10^5$ 
samples.
For $d = 2$ the network evaluates $\ptri$ in $0.04\,\mu$s with root-mean-square 
error less than $10^{-3}$ (see Section \ref{sec:nn_h}). 
By contrast, Monte Carlo with $10^5$ samples takes 
roughly $1{,}300\,\mu$s for comparable error.
The algorithms we implemented use $\rho$ of \eqref{eq:dist} with 
$S = \tfrac{2}{3} \Sx + \tfrac{1}{3} \Sy$, as in \cite{hong2026}, though all
the analysis follows through with any convex combination of $\Sx, \Sy$.
We use our algorithm for evaluating $\ptri$ to find the MAP estimate of the color 
discrimination model of \cite{hong2026}. Our algorithm produces a MAP estimate  
with covariance field relative error $0.8\%$ in $0.30$\,seconds,
compared to the Monte Carlo approach which produces relative error
$2.9\%$ after $44$ minutes (Section \ref{sec:triangle}).

For the duo-trio test with $\Sx = \Sy $, the function $\pduo$ depends on the stimuli and
covariance only through the scalar $r = (\my-\mx)^\top \Sx^{-1} (\my-\mx)$. We
approximate $\pduo$, after the transformation of variables, with a Chebyshev expansion in 
$r$ whose degree is set by a user-specified accuracy $\epsilon$. 
The coefficients of the Chebyshev expansion were computed by using an 
adaptive quadrature algorithm to evaluate $\pduo$ via a $1$-dimensional integral 
obtained after the transformation of variables. 
For $d = 3$ and $\epsilon = 10^{-8}$, an expansion with $36$ terms evaluates
$\pduo$ with error less than $10^{-11}$ in $0.028\,\mu$s (Section \ref{sec:duotrio}).

The two tests we consider --- triangle and duo-trio --- and the option to assume equal 
covariances $\Sx = \Sy$
represent four typical scenarios that appear in sensory discrimination models. 
We address three of them --- the triangle test with equal or unequal covariances
(Section \ref{sec:triangle}) and the duo-trio test with equal covariances
(Section \ref{sec:duotrio}).
For the fourth case, the duo-trio test with unequal covariances, analysis and efficient algorithms 
have been introduced~\cite{mullen1991, ennis1994}. 
Although we do not treat it, that case is also amenable to the approach 
developed here for $\ptri$ and $\pduo$. 

The remainder of the paper is organized as follows. 
Section \ref{sec:math_app} introduces theoretical tools that are used throughout the remainder 
of the paper, including the affine invariance of $\ptri, \pduo$. 
Sections \ref{sec:triangle} and \ref{sec:duotrio} describe fast algorithms for efficiently evaluating
$\ptri, \pduo$. In Section \ref{sec:discussion}, we discuss extensions and limitations. Finally, 
Appendix \ref{sec:ml_rule} relates the choice of distance function to that of an ``ideal observer'' and makes rigorous the notion of ``ideal observer''. 

The algorithms of this paper ---  the neural networks of Section \ref{sec:nn_h} and 
the Chebyshev expansions of Section \ref{sec:duotrio} --- were implemented in Python and
are publicly available at \url{https://github.com/pgree/fast-likelihood-sensory-discrimination}.
That repository also includes code for generating the figures of this paper. 

\section{Mathematical apparatus}\label{sec:math_app}
The primary purpose of this section is Claim \ref{claim:linear_invariance}, which is used 
to reduce the dimensionality of the arguments of $\ptri$ and $\pduo$. First, we introduce several supporting 
analytical results that will be used in the proof of Claim \ref{claim:linear_invariance}. 
Throughout this section, we denote by $\phi(x; \mu, \Sigma)$ the Gaussian p.d.f., 
\begin{align}
\phi(x; \mu, \Sigma) = \frac{1}{(2\pi)^{d/2} |\Sigma|^{1/2}}
e^{-\tfrac{1}{2} (x - \mu)^\top \Sigma^{-1} (x - \mu)}.
\end{align}
The following lemma is a standard identity for a linear change of variables in a Gaussian probability 
density function. 
\begin{lemma}\label{lem:change_of_v}
For any invertible $d \times d$ matrix $A$, any $\mu, b \in \R^d$, and any positive definite matrix 
$\Sigma \in \R^{d \times d}$, we have
\begin{align}
\phi(x; \mu, \Sigma) = |\det A|\, \phi(Ax + b;\, A \mu + b,\, A \Sigma A^\top).
\end{align}
\end{lemma}
In the following, we provide an identity for $\rho_{X, Y}$ under the same affine transformation 
of $X \sim \mcN(\mx, \Sx), Y\sim \mcN(\my, \Sy)$, where $\rho_{X, Y}$ denotes the distance \eqref{eq:dist} with $S = t \Sx + (1 - t) \Sy$ for any $t \in [0, 1]$. 
\begin{lemma}\label{lem:dist_inv}
For any invertible $d \times d$ matrix $A$ and any $b \in \R^d$,
\begin{align}
\rho_{AX + b, AY + b}(Ax + b, Ay + b) = \rho_{X, Y}(x, y).
\end{align}
\end{lemma}
\begin{proof}
The matrix $S$ (see \eqref{eq:dist}) for the transformed Gaussians $AX + b, AY + b$ is
given by $ASA^\top = t\, A \Sx A^\top + (1 - t)\, A \Sy A^\top$. Since
$(Ax + b) - (Ay + b) = A(x-y)$, we have
\begin{equation}
\begin{aligned}
\rho_{AX + b, AY + b}(Ax + b, Ay + b)
&= (Ax - Ay)^\top (A S A^\top)^{-1} (Ax - Ay) \\
&= (x - y)^\top A^\top A^{-\top} S^{-1} A^{-1} A\, (x - y) \\
&= (x - y)^\top S^{-1} (x - y) \\
&= \rho_{X, Y}(x, y).
\end{aligned}
\end{equation}
\end{proof}
The following corollary, which follows immediately from Lemma \ref{lem:dist_inv}, is used in Claim \ref{claim:linear_invariance}. 
\begin{corollary}\label{lem:set_pullback}
Let $X \sim \mcN(\mx, \Sx)$ and $Y \sim \mcN(\my, \Sy)$. 
Let $S^{\mathrm{tri}}_{X,Y}$ and
$S^{\mathrm{duo}}_{X,Y}$ denote the sets $S_{\mathrm{tri}}$ and $S_{\mathrm{duo}}$
of \eqref{eq:S_tri}, \eqref{eq:S_duo} with $\rho = \rho_{X,Y}$.
For any invertible $d \times d$ matrix $A$ and any $b \in \R^d$, we have
\begin{align}
S^{\mathrm{tri}}_{AX + b, AY + b} &= \{ (A x_1 + b, A x_2 + b, A y + b) : (x_1, x_2, y) \in S^{\mathrm{tri}}_{X, Y} \}, \\
S^{\mathrm{duo}}_{AX + b, AY + b} &= \{ (A x_{\mathrm{ref}} + b, A x_1 + b, A y_1 + b) :
   (x_{\mathrm{ref}}, x_1, y_1) \in S^{\mathrm{duo}}_{X, Y} \}.
\end{align}
\end{corollary}
The following, which establishes the affine invariance of $\ptri, \pduo$ 
is the main analytical result of this section. 
\begin{claim}\label{claim:linear_invariance}
For any invertible $d \times d$ matrix $A$ and any $b \in \R^d$, the functions $\ptri$ and
$\pduo$ (see~\eqref{eq:p_triangle}, \eqref{eq:p_duotrio}) are preserved when the inputted Gaussians are
each transformed by the affine transformation $x \mapsto Ax + b$. That is,
\begin{align}\label{eq:f_equiv}
  \ptri(\mx, \my, \Sx, \Sy)
  &= \ptri(A\mx + b,\, A\my + b,\, A \Sx A^\top,\, A \Sy A^\top), \\
  \pduo(\mx, \my, \Sx, \Sy) &= \pduo(A\mx + b,\, A\my + b,\, A \Sx A^\top,\, A \Sy A^\top).
  \label{eq:g_equiv}
\end{align}
\end{claim}
\begin{proof}
We begin by proving equation \eqref{eq:f_equiv}. 
Substituting Lemma \ref{lem:change_of_v} into \eqref{eq:f_tri_int},
we have 
\begin{equation}\label{eq:ptri_mid}
\begin{aligned}
  \ptri(\mx, \my, \Sx, \Sy)
  = \int_{S^{\mathrm{tri}}_{X, Y}}
      \phi(A x_1 + &b; A\mx + b, A\Sx A^\top)\,\phi(A x_2 + b; A\mx + b, A\Sx A^\top)\, \\
      &\phi(A y + b; A\my + b, A\Sy A^\top)\,
      |\det A|^{3}\,\mathrm{d}x_1\, \mathrm{d}x_2\, \mathrm{d}y.
\end{aligned}
\end{equation}
Applying the change of variables $\tx_1 = A x_1 + b$, $\tx_2 = A x_2 + b$, and
$\ty = A y + b$ transforms the 
differentials via $\mathrm{d}\tx_i = |\det A|\,\mathrm{d}x_i$, $\mathrm{d}\ty = |\det A|\,\mathrm{d}y$,
and the domain of integration becomes $S^{\mathrm{tri}}_{AX + b, AY + b}$ by Corollary \ref{lem:set_pullback}. That is,
\begin{equation}\label{eq:f_last_step}
\begin{aligned}
  \ptri(\mx, \my, \Sx, \Sy) 
 = \int_{S^{\mathrm{tri}}_{AX + b, AY + b}}
      \phi(\tx_1; A\mx + b, & A\Sx A^\top)\,
      \phi(\tx_2; A\mx + b, A\Sx A^\top)\, \\
      & \phi(\ty; A\my + b, A\Sy A^\top)\,
      \,\mathrm{d}\tx_1\, \mathrm{d}\tx_2\, \mathrm{d}\ty.
\end{aligned}
\end{equation}
Substituting \eqref{eq:f_tri_int} into \eqref{eq:f_last_step}, we obtain \eqref{eq:f_equiv}.

Identity \eqref{eq:g_equiv} is obtained by the same computation, with the
integrand of \eqref{eq:f_tri_int} replaced by that of \eqref{eq:f_duo_int}. Using Corollary \ref{lem:set_pullback},
the domain of integration of \eqref{eq:ptri_mid}, $S^{\mathrm{tri}}_{X, Y}$, is replaced by $S^{\mathrm{duo}}_{X, Y}$, 
whose image under $x \mapsto Ax + b$ is $S^{\mathrm{duo}}_{AX + b, AY + b}$.
\end{proof}

The following is a technical lemma used in the proof of Claim \ref{claim:duotrio_mahal} to 
demonstrate the duo-trio probability \eqref{eq:f_duo_int} can be reduced to a function of one variable. 
\begin{lemma}\label{lem:duotrio}
Let $x_0, x_1 \in \R^d$ be independent Gaussians $\mcN(0, I)$ and let 
$y \sim \mcN(\mu, I)$. Define $\mbfu=[u_0\,\, u_1]$
by 
\begin{align}
u_0 = x_1-x_0, \qquad u_1 = y - x_0. 
\end{align}
Then 
\begin{align}
  \mbfu \sim \mcN \!\left(\begin{bmatrix}0\\\mu\end{bmatrix},\, C \right),
  \qquad C = 
  \begin{bmatrix}2I & I\\ I & 2I\end{bmatrix}.
\end{align}
Let $\mbfz = [z_0 \,\, z_1] $ where $z_0, z_1 \in \R^d$ are defined by 
$\mbfu = C^{1/2} \mbfz$. Then 
$\mbfz = [z_0 \,\, z_1] \sim \mcN(m, I)$ where 
\begin{align}\label{eq:m_def}
m = C^{-1/2} [0 \,\, \mu]^\top = \bigg[\tfrac{\sqrt3-3}{6}\mu \,\,\, \tfrac{3+\sqrt3}{6} \mu\bigg]^\top.
\end{align}
\end{lemma}
\begin{proof}
Since $\mbfu$ is a linear combination of Gaussians 
$x_0, x_1, y$, we know that $\mbfu$ is also Gaussian. By linearity 
of expectation, $\E [u_0] = 0$ and $\E [u_1] = \mu$. Since $x_0, x_1, y$ are independent,
\begin{equation}
  \var(u_0) = \var(u_1) = 2I, \qquad \cov(u_0, u_1) = \var(x_0) = I.
  \qquad
\end{equation}
Straightforward calculation shows that 
\begin{equation}\label{eq:c12_def}
  C^{1/2} = \begin{bmatrix}\alpha I & \beta I\\ \beta I & \alpha I\end{bmatrix},
  \qquad
  \alpha = \frac{\sqrt3+1}{2},\quad \beta = \frac{\sqrt3-1}{2}
\end{equation}
and thus 
\begin{equation}\label{eq:cm12_def}
  C^{-1/2}
  = \frac{1}{\alpha^2-\beta^2}\begin{bmatrix}\alpha I & -\beta I\\ -\beta I & \alpha I\end{bmatrix}
  = \begin{bmatrix}
      \tfrac{3+\sqrt3}{6} I & \tfrac{\sqrt3-3}{6} I\\[2pt]
      \tfrac{\sqrt3-3}{6} I & \tfrac{3+\sqrt3}{6} I
    \end{bmatrix}.
\end{equation}
Equation \eqref{eq:m_def} follows immediately from equation \eqref{eq:cm12_def} and 
the fact that 
\begin{align}
\E[z] = \E[C^{-1/2} u] = C^{-1/2} [0 \,\, \mu]^\top.
\end{align}
\end{proof}

\section{Triangle probability}\label{sec:triangle}
In this section we introduce an efficient algorithm for approximating the
function $\ptri$ (see \eqref{eq:p_triangle}) that decomposes $\ptri$ into $\ptri = h \circ g$
where $g: \R^{2d + d(d + 1)} \to \R^{2d}$ is computationally inexpensive to evaluate
and $h : \R^{2d} \to [0, 1]$ is approximated with a small neural network.
Throughout this section, $\rho$ is \eqref{eq:dist} with $S = \tfrac{2}{3}\Sx + \tfrac{1}{3}\Sy$.

\subsection{Reducing $\ptri$ to $2d$ arguments}\label{sec:reduction}
The primary mathematical tool for reducing the dimensionality of $\ptri$ is Claim \ref{claim:linear_invariance}
and the transformation follows two steps. 
First, both Gaussians of the Thurstonian model are translated so that one has mean at the origin. Then, 
the covariance matrices of the Gaussians are simultaneously diagonalized.
Here, we describe the two steps involved in reducing $\ptri$ from $2d + d(d + 1)$ arguments to $2d$
arguments.

First, $\ptri(\mx, \my, \Sx, \Sy)$ is preserved by translating both Gaussians, or 
adding the same vector to $\mx$ and $\my$. In particular, 
using Claim \ref{claim:linear_invariance}, we have
\begin{align}\label{eq:ptri0}
\ptri(\mx, \my, \Sx, \Sy) = \ptri(0, \my - \mx, \Sx, \Sy)
\end{align}
for all $\mx, \my \in \R^d$ and all $\Sx, \Sy$. 

We next simultaneously diagonalize $S$ of \eqref{eq:dist} and $\Sx$. Since $S$ is a
linear combination of $\Sx$ and $\Sy$, simultaneously diagonalizing any two of $S$, $\Sx$, $\Sy$ diagonalizes 
the third.
Specifically, we start by transforming $S$ (see \eqref{eq:dist}) to the identity matrix. This transformation 
results in a Euclidean distance function $\rho$ of \eqref{eq:dist}.
That is, we multiply the Gaussians\footnote{In this section, for expositional convenience, we refer to $\ptri$ as a function of two Gaussians. That is, $\ptri(\mx, \my, \Sx, \Sy)$ is a function of the Gaussians $\mcN(\mx, \Sx)$ and $\mcN(\my, \Sy)$.}
of the right hand side of \eqref{eq:ptri0}
by the matrix 
$M = L^{-1}$, where $S = L L^\top$ is the Cholesky factorization of $S$,
so that $M S M^\top = I$. Note that any $M$ satisfying $M S M^\top = I$ is sufficient.
Using Claim \ref{claim:linear_invariance}, we know that this transformation preserves 
$\ptri$. That is, 
\begin{align}\label{eq:ptri1}
\ptri(\mx, \my, \Sx, \Sy) = \ptri(0, M(\my - \mx), M \Sx M^\top , M \Sy M^\top).
\end{align}
We next transform the remaining Gaussian arguments of $\ptri$ 
so that their covariances are diagonal. 
Using the eigendecomposition
\begin{align}\label{eq:M}
M \Sx M^\top = V \Lambda V^\top,
\end{align}
we transform the Gaussians of the right hand side of \eqref{eq:ptri1} via
\begin{align}\label{eq:ptri2}
\ptri(\mx, \my, \Sx, \Sy) = \ptri(0, V^\top M(\my - \mx), V^\top M \Sx M^\top V  , V^\top M \Sy M^\top V).
\end{align}
The covariances of \eqref{eq:ptri2} are related via the identities
\begin{align}\label{eq:cov_ids}
\tfrac{2}{3} \Stx+ \tfrac{1}{3} \Sty = I, \qquad \Stx = V^\top M \Sx M^\top V, \qquad \Sty = V^\top M \Sy M^\top V
\end{align}
and both $\Stx, \Sty$ are diagonal. All together, applying the 
affine transformation 
\begin{align}
x \mapsto V^\top M (x - \mx),
\end{align}
to the Gaussians of $\ptri(\mx, \my, \Sx, \Sy)$, and using identity \eqref{eq:cov_ids}, we have 
\begin{align}\label{eq:ptri3}
\ptri(\mx, \my, \Sx, \Sy) = \ptri(0, V^\top M(\my - \mx), \Stx, 3I - 2\Stx)
\end{align}
where $\Stx$ is a diagonal covariance matrix. 

\begin{remark}\label{rem:mu_sign}
The eigendecomposition \eqref{eq:M}
is only determined up to the sign of the eigenvectors (columns of $V$), provided that 
there are no repeated eigenvalues, which occurs on a set of measure zero.
Multiplying column $j$ of $V$ by $-1$ leaves $\Stx, \Sty$ unchanged, but does change
the sign of component $j$ of the argument $V^\top M (\my - \mx)$ of \eqref{eq:ptri3}.
Thus, we arbitrarily choose $V$ such that each component of the mean is non-negative.
Equivalently, we can transform the vector $V^\top M (\my - \mx)$ by taking the absolute 
value of each of its components.
\end{remark}
\begin{remark}\label{rem:eig_order}
We order the diagonal entries $\lambda_1, \ldots, \lambda_d$ of $\Lambda$, and accordingly
the columns of $V$, so that $\lambda_1 \geq \cdots \geq \lambda_d$.
For any permutation matrix $P$,  the matrix $P \Lambda P^\top$ is the diagonal
matrix of the permuted eigenvalues, so by Claim \ref{claim:linear_invariance}, permuting 
the entries of $\Lambda$ and, accordingly, the columns of $V$, leaves $\ptri$ unchanged.
\end{remark}

After the above transformation, $\ptri$ becomes a function of the $2d$ arguments 
$(\lambda, \eta)$, where $\lambda \in \R^d$ is the diagonal of $\Lambda$ and 
we define $\eta$ by $\eta = |V^\top M (\my - \mx)| \in \R_{\geq 0}^d$ 
where the absolute value is taken element-wise. 
For notational convenience, we denote by $g: \R^{2d + d(d+1)} \to \R^{2d}$
the function that maps $(\mx, \my, \Sx, \Sy)$  to $(\lambda, \eta)$ and 
we define $h$ to be the function 
\begin{align}\label{eq:g_h_def}
h(\lambda, \eta)  = h(g(\mx, \my, \Sx, \Sy)) = \ptri(\mx, \my, \Sx, \Sy).
\end{align}

In the following, we demonstrate that after the transformation by $g$ of the arguments 
of $\ptri$, the function $h$ has arguments
$\lambda \in (0, \tfrac{3}{2})^d$ and $\eta_i \geq 0$ for $i=1,\dots,d$. 
\begin{claim}\label{claim:domain}
Let 
\begin{align}
D_d = \{(\mx, \my, \Sx, \Sy) : \mx, \my \in \R^d, \text{ and symmetric positive definite } 
\Sx, \Sy \in \R^{d \times d} \}\nonumber
\end{align}
and let 
\begin{align}\label{eq:domain}
  \Omega_d = \bigl\{ \lambda \in \R^d : \tfrac{3}{2} > \lambda_1 \geq \cdots \geq \lambda_d > 0 \bigr\} \times \R_{\geq 0}^d .
\end{align}
Then $g$ of \eqref{eq:g_h_def} maps $D_d$ onto $\Omega_d$.
\end{claim}
\begin{proof}
We first demonstrate that $g(D_d) \subseteq \Omega_d$.
The vector $\lambda$ of \eqref{eq:g_h_def} is the diagonal of $\Stx$, 
and $\Stx = (3I - \Sty)/2$ by \eqref{eq:cov_ids}. 
Positive definiteness of $\Stx$ implies $\lambda_i > 0$ and positive definiteness of 
$\Sty$ gives $\lambda_i < 3/2$ for $i=1,\dots, d$. 
Since $\eta_i \geq 0$ and $\lambda_1 \geq \cdots \geq \lambda_d$ by 
Remarks \ref{rem:mu_sign} and \ref{rem:eig_order}, we have $g(D_d) \subseteq \Omega_d$. 

Now we demonstrate that $\Omega_d \subseteq g(D_d)$. Let 
$(\lambda, \eta) \in \Omega_d$ and set $\mx = 0$, $\my = \eta$, $\Sx = \Lambda$ 
and $\Sy = 3I - 2 \Lambda$. 
Then %
$g(\mx, \my, \Sx, \Sy) = (\lambda, \eta)$.
\end{proof}

\subsection{Neural network approximation of $h$}\label{sec:nn_h}
After the transformation by $g$ described in the 
previous section, we can now evaluate $\ptri$ by evaluating the function 
$h: \Omega_d \to [0, 1]$. That is, we have now reduced the problem of efficiently
evaluating $\ptri$, a function of $2d + d(d+1)$ arguments, to the problem
of evaluating $h$, a function of $2d$ arguments. 
Unfortunately, for stimulus spaces in more than one dimension, $2d$ dimensions is
still generally too large to evaluate $h$ efficiently with standard tensor-product 
quadrature rules, whose cost scales as $\mathcal{O}(n^{2d})$ where $n$ is the number 
of nodes in one dimension. 
To avoid this exponential computational cost, we approximate $h$ with a small neural 
network. It turns out that small, classical, neural networks are able to approximate 
$h$ to reasonable accuracy and that after training, these neural networks can 
be evaluated orders of magnitude faster than Monte Carlo for comparable accuracy. 

We train the neural network with labeled data calculated through 
Monte Carlo evaluation of $\ptri$. 
We choose training points in $\Omega_d$ by the following procedure.
We first sample $\lambda_i$ uniformly over $[10^{-3}, \tfrac{3}{2} - 10^{-3}]^d$ and then order
the points.\footnote{As $\lambda_i \to 0$, the condition number of $\Sx$ approaches
infinity, and as $\lambda_i \to \tfrac{3}{2}$ the condition number of $\Sy$ 
approaches infinity. These are regimes that are generally not relevant to the applications 
we consider. We thus restrict $\lambda_i \in [\epsilon, \tfrac{3}{2} - \epsilon]$ with 
$\epsilon = 10^{-3}$.}
We then sample $\eta_i$ by first uniformly sampling the 
direction $\eta / \| \eta \|$ on the unit sphere restricted to the positive orthant.  
The norm $\| \eta \|$ is then uniformly sampled on $(0, R)$ for
$R > 0$ chosen as follows. We choose $R$ such that $h(\lambda, \eta)$ is
well approximated by $1$ for all $\lambda_i \in (0, \tfrac{3}{2})$ and $\| \eta \| > R$. 

We use the following two observations to construct a numerical procedure to 
choose $R$. 
First, we observe empirically that $h$ is an increasing function in each 
$\eta_i$ for fixed $\lambda$. Additionally, for fixed $\| \eta \|$, the maximum of
$1 - h(\lambda, \eta)$ over $\eta$ and $\lambda \in [\gamma, \tfrac{3}{2} - \gamma]^d$ 
is attained at $\lambda = (\tfrac{3}{2} - \gamma)(1, \dots, 1)$ for small $\gamma > 0$.
Next, we observe that for $\lambda$ with constant entries, $h(\lambda, \eta)$ is 
invariant to orthogonal transformations of $\eta$. 

Using these two observations, we choose $R$ by the following mechanism. 
We set $\lambda = (\tfrac{3}{2} - 10^{-3}, \dots, \tfrac{3}{2} - 10^{-3})$ and 
$\eta = \| \eta \| (1, 0, \dots, 0)$, where the angular component of $\eta$
is arbitrarily chosen since for $\lambda$ with constant entries, 
$h(\lambda, \eta)$ depends on $\eta$ only 
through $\| \eta \|$. We then evaluate
$h(\lambda, \eta)$ via Monte Carlo on a $1$-dimensional grid in $\| \eta \|$.
We compute the smallest $\| \eta \| $ such that
$1 - h(\lambda, \eta) < \delta$ for some error tolerance $\delta$. We find
$\| \eta \| = 10.9$ with $\delta = 10^{-4}$ for $d=2$, and $\| \eta \| = 11.1$ for $d = 3$, and
set $R = 12$ for both. 
We prove that the network output for $d=2$ is within $2.2 \times 10^{-5}$ of $1$ for all
$\| \eta \| \geq R$ and $\lambda_i \in [10^{-3}, \tfrac{3}{2} - 10^{-3}]$.\footnote{The proof uses the formal verification software Super~\cite{super2026} and relies on the combination of two results. One gives the bound for $R \leq \| \eta \| \leq 1000$. A second shows that the logit of the network output is at least $\| \eta \| - 323$, so that the bound is satisfied for $\| \eta \| > 1000$.
Both results are in the repository \url{https://github.com/pgree/fast-likelihood-sensory-discrimination}.}

We evaluate $\ptri$ for training data with $10^5$ Monte Carlo samples. 
The training set we use consists of $3 \times 10^5$ data points. 
Each of the $3 \times 10^5$ training labels has error with standard deviation
$\sqrt{h(1 - h)/n}$ where $h$ is the true value and $n$ is the number of Monte 
Carlo samples. Thus the maximum standard deviation is bounded above by $0.002$. 
With $\| \eta \|$ uniform on $(0, R)$, the training set has a higher concentration of 
points near the origin than near $R$. In particular, the density per unit volume is 
proportional to $\| \eta \|^{1 - d}$.
We make this choice because when $\| \eta \|$ is near $R$, the function $\ptri$ is 
close to $1$. We observe empirically that with this choice of training distribution, 
the error on a validation set is smaller than for $\eta$ sampled uniformly over 
the ball $\| \eta \| \leq R$.

The network architecture we use for $d=2, 3$ is a multilayer perceptron with layer widths
$2d$, $16$, $16$, $1$, GELU activation functions, and a logistic output unit. 
The inputs to the network are $(\log \lambda, \eta)$, standardized to zero mean 
and unit variance. We observe empirically 
that when using an input of $\lambda$, instead of $\log \lambda$, large errors 
concentrate near $\lambda = 0$. 
The network has $369$ parameters for $d = 2$ and $401$ for $d = 3$. 
We train the network by minimizing cross-entropy with Adam~\cite{kingma2015}, batch size 
$1024$, for $1.12 \times 10^5$ steps with the learning rate decayed from 
$10^{-3}$ on a cosine schedule, and retain the best of three restarts as scored
by the empirical risk on the training labels.

We measure the accuracy of the network as follows. We evaluate it at $4992$ points 
not used in training, with $\lambda_i \in [10^{-3}, \tfrac{3}{2} - 10^{-3}]$ and
$\| \eta \| \leq 12$. 
The test set consists of $4992$ points chosen so that $h$ evaluated at the test set 
is roughly uniform on $[\tfrac{1}{3}, 1]$. 
We start by drawing $12 \times 4992$ candidates from the distribution of the training
points. We then divide
$[\tfrac{1}{3}, 1]$ into $24$ intervals of equal length, and draw $208$ samples from 
each interval without replacement, giving
$24 \times 208 = 4992$ points. 
At each point in the test set we compute $h$ by Monte Carlo with $n = 10^7$ samples. 
The standard deviation
of these reference values is at most $1.6 \times 10^{-4}$. We find that the root-mean-square 
difference between the network and the reference values is $7.9 \times 10^{-4}$ for $d = 2$ and
$9.3 \times 10^{-4}$ for $d = 3$, with standard errors $1.7 \times 10^{-5}$ and $1.8 \times 10^{-5}$, 
and the maximum error over the $4992$ points is $6.6 \times 10^{-3}$ and $7.0 \times 10^{-3}$. 
The training labels, computed with $n = 10^5$
samples, have root-mean-square standard deviation $8.3 \times 10^{-4}$, thus the error of the
network is comparable to the accuracy with which its training labels were measured.

\subsection{Neural network on the original arguments}
In this section, we have thus far described a method to approximate $\ptri$ with the 
composition of functions $h \circ g$ where $g$ performs the dimensionality reduction 
and $h$ is approximated with a neural network. 
A natural alternative to this approach is to approximate $\ptri$ with a network whose inputs 
are the $d + d(d + 1)$ arguments of \eqref{eq:p_triangle}
(removing $d$ arguments from $\ptri$ through translation invariance of the means). 

\begin{figure}[h]
\centering
\begin{subfigure}[t]{0.49\textwidth}
\centering
\includegraphics[width=\textwidth]{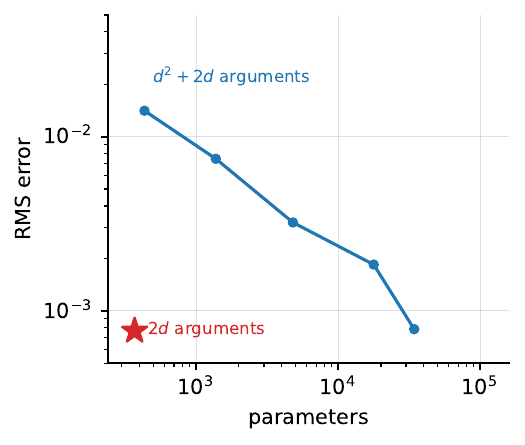}
\caption{number of parameters}
\label{fig:reduction_params}
\end{subfigure}
\hfill
\begin{subfigure}[t]{0.49\textwidth}
\centering
\includegraphics[width=\textwidth]{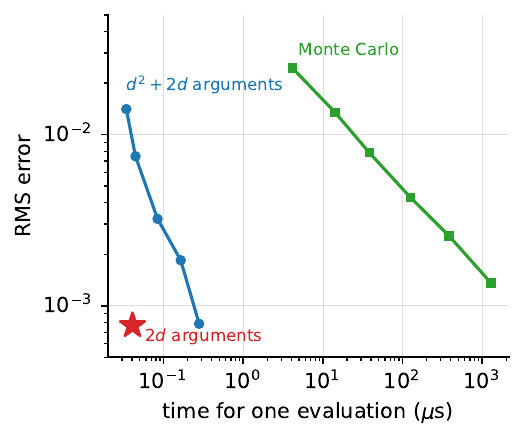}
\caption{time for one evaluation}
\label{fig:reduction_time}
\end{subfigure}
\caption{Accuracy for a single evaluation of $\ptri$ as a function of
number of parameters (Figure \ref{fig:reduction_params}) and time for evaluation (Figure \ref{fig:reduction_time}).
We compare the neural networks of various sizes trained on $\ptri$ with the original
number of arguments (blue) and the neural network trained on the $2d$ inputs of
$h$ (see \eqref{eq:g_h_def}) (red). In Figure \ref{fig:reduction_time}, the time for evaluation, we compare 
the two neural network approaches to the Monte Carlo approach of \cite{hong2026} (green).
Errors are root-mean-square differences, for $d = 2$, over a set of $4992$ points
whose reference values are computed by Monte Carlo with $n = 10^7$ samples.}
\label{fig:reduction}
\end{figure}

We trained neural networks of various sizes on $\ptri$ with its original $d + d(d + 1)$ 
arguments and compare performance with (i) a neural network trained on $h$ 
after dimensionality reduction and (ii) Monte Carlo. 
In Figure \ref{fig:reduction}, we compare the two neural networks in terms of accuracy 
as a function of number of parameters in the network (Figure \ref{fig:reduction_params})
and in Figure \ref{fig:reduction_time} we compare the two networks and Monte Carlo
in accuracy versus time for one evaluation. 
Figure \ref{fig:reduction} uses $d=2$ and errors are root-mean-square
differences over a validation set of $4992$ points, sampled and labelled in the same way 
as the test set. 

For the larger network, we trained networks of width $16$, $32$, $64$ and 
$128$ at depth $2$ and of width $128$ at depth $3$.
The larger network required two orders of magnitude more parameters to achieve 
the same accuracy, and about seven times more time for evaluation. 
We note that all networks with the same number of arguments were trained on the 
same training data. 
The training data we use for the large networks (those trained on the original $2d + d^2$
arguments) is obtained from a transformation of the training data used for the small 
networks (those trained to approximate $h$).
Specifically, for $(\lambda, \eta) \in \Omega_d$, the preimage
$g^{-1}(\lambda, \eta) = \{ (\mx, \my, \Sx, \Sy) \in D_d : g(\mx, \my, \Sx, \Sy) = (\lambda, \eta) \}$
is nonempty by Claim \ref{claim:domain}, and we choose an element of it randomly as follows.
We set $A = Q D_s$, where $Q$ is a rotation drawn uniformly from $SO(d)$ and
the $\log_{10}$ of each diagonal element of the diagonal matrix $D_s$ is 
independent and uniform on $(-\tfrac{1}{4}, \tfrac{1}{4})$. We then set
$\mx = 0$, $\my = A \eta$, $\Sx = A \diag(\lambda) A^\top$ and $\Sy = A (3I - 2 \diag(\lambda)) A^\top$.
\begin{remark}
The construction of the training set for the neural network trained on the original arguements
relies on the reduction of variables. The training points are drawn from the
$2d$-dimensional set $\Omega_d$ rather than from the $(d + d(d+1))$-dimensional space of the
original arguments. A tensor-product grid with $n$ points in each dimension would have
$n^{d + d(d+1)}$ points instead of $n^{2d}$.
\end{remark}

Timings are wall-clock times on the CPU of an Apple M3 Max laptop,
using JAX 0.10.0 \cite{jax2018}, and exclude compilation. Networks are timed on batches of
$20{,}000$ trials.
For $d = 2$, the function $g$ that reduces the dimensionality of $\ptri$ and the network 
take $0.041\,\mu$s per evaluation provided that the eigendecomposition \eqref{eq:M} is computed 
in closed form. At the same accuracy, 
the network on the original arguments takes $0.28\,\mu$s.

\subsection{Fast algorithm for statistical inference} 
The color discrimination model \cite{hong2026} uses an experimental procedure 
involving a triangle test (see Figure \ref{fig:trials} for an example) in a two-dimensional
stimulus space, the so-called iso-luminant plane. A Bayesian model is fit to the 
experimental trial data. The experiment consists of $N = 6000$ trials from participant CH of \cite{hong2026}. In each trial
$i = 1, \dots, N$, the subject is shown the repeated stimulus $\mu_{\mathrm{x},i}$ twice and the unique
stimulus $\mu_{\mathrm{y},i}$ once, and $y_i = 1$ if the response is correct and $y_i = 0$ otherwise.
The model is
\begin{align}\label{eq:color_model}
y_i \mid \theta &\sim \mathrm{Bernoulli}\bigl( p_i(\theta) \bigr), \\
p_i(\theta) & = \ptri\bigl(\mu_{\mathrm{x},i}, \mu_{\mathrm{y},i}, \Sigma(\mu_{\mathrm{x},i}, \theta), \Sigma(\mu_{\mathrm{y},i}, \theta)\bigr), \\
\Sigma(x, \theta) &= U(x, \theta)\, U(x, \theta)^\top, \\
U_{ab}(x, \theta) & = \sum_{k, \ell = 0}^{4} \theta_{k \ell a b}\, T_k(\xi_1)\, T_\ell(\xi_2), \nonumber \\
\theta_{k \ell a b} &\sim \mcN\bigl(0,\, 3 \times 10^{-4} \cdot 0.4^{k + \ell}\bigr). \nonumber
\end{align}
Here $T_k$ is the Chebyshev polynomial of degree $k$, $x = (\xi_1, \xi_2) \in [-1, 1]^2$,
$a \in \{1, 2\}$ and $b \in \{1, 2, 3\}$, so that $U(x, \theta)$ is a
$2 \times 3$ matrix. The $y_i$ are independent given $\theta$, and the $\theta_{k \ell a b}$ are
independent.
For inference, \cite{hong2026} uses a MAP estimate. That is, they solve for
\begin{align}\label{eq:map}
\htheta = \arg\max_{\theta} \Bigl( \sum_{i=1}^N \bigl[ y_i \log p_i(\theta) + (1 - y_i) \log\bigl(1 - p_i(\theta)\bigr) \bigr]
  + \log p(\theta) \Bigr),
\end{align}
where $p(\theta)$ is the prior density. They use a Monte Carlo estimate for the triangle 
probability $\ptri$.

\begin{figure}[h]
\centering
\begin{subfigure}[t]{0.49\textwidth}
\centering
\includegraphics[width=\textwidth]{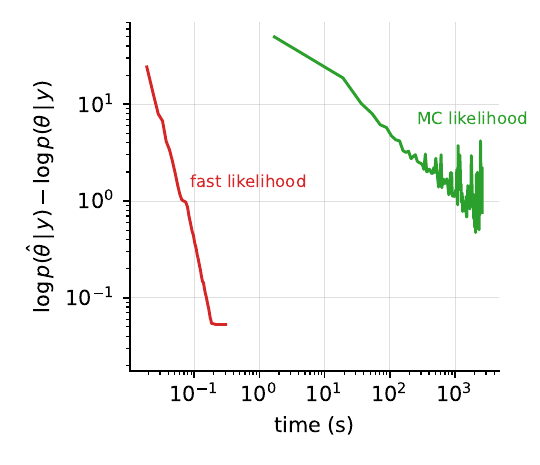}
\caption{objective function discrepancy}
\label{fig:map_timing_gap}
\end{subfigure}
\hfill
\begin{subfigure}[t]{0.49\textwidth}
\centering
\includegraphics[width=\textwidth]{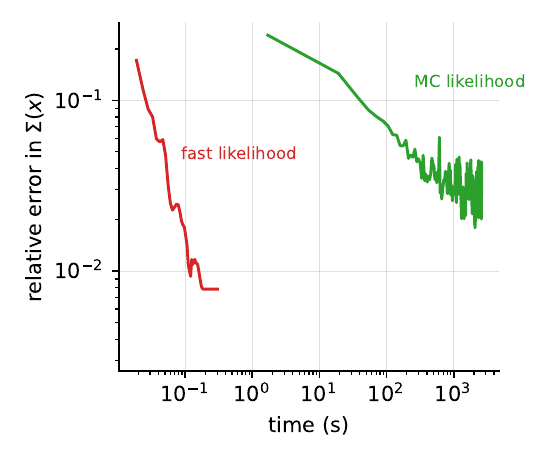}
\caption{covariance field discrepancy}
\label{fig:map_timing_sigma}
\end{subfigure}
\caption{Comparison of time versus accuracy for the MAP estimate of 
the model in \cite{hong2026}, where 
$\ptri$ is evaluated by the Monte Carlo likelihood (green) and by the fast
likelihood (red). We show (a) the log posterior at the MAP, $\htheta$, minus the 
log posterior at the current $\theta$, and (b) the relative error in 
$\Sigma(x)$ at the current parameters $\theta$ compared to $\Sigma(x)$
at the MAP. 
The MAP $\htheta$ is the best of $200$ L-BFGS-B fits with the fast likelihood. 
Monte Carlo with $n = 10^5$ gives the same final gaps in (a) to within $0.09$.}
\label{fig:map_timing_local}
\end{figure}

Here, we use the fast likelihood approximation of $\ptri$ for fitting the same model
also with a MAP estimate. 
We compare timing and accuracy of the two approaches on a CPU.
Figure \ref{fig:map_timing_local} provides two comparisons of time as a function of accuracy. 
Figure \ref{fig:map_timing_gap} illustrates $\log p(\htheta | y) - \log p(\theta | y)$ as a function
of time, where $p(\theta | y)$ is the posterior with $\ptri$ evaluated by the fast likelihood.
Figure \ref{fig:map_timing_sigma} demonstrates the relative 
error 
\begin{align}
\left( \frac{1}{m} \sum_{j=1}^m
\frac{\| \Sigma(x_j, \theta) - \Sigma(x_j, \htheta) \|_F^2}{\| \Sigma(x_j, \htheta) \|_F^2}
\right)^{1/2}
\end{align}
in the approximation of the covariance field at a set of points $x_1,\dots,x_m$ on a 
Cartesian grid. Here, $\theta$ is the current iterate of each method and $\htheta$ is the
best known MAP of the posterior with the fast likelihood, the best of $200$ L-BFGS-B fits from independent prior draws. Both fits start from $\htheta + 0.2\,\xi$, with $\xi$ a prior draw. The curves in Figure \ref{fig:map_timing_local} begin after the 
first iteration of each method. With the fast likelihood, L-BFGS-B ends with $\log p(\htheta | y) - \log p(\theta | y) = 0.053$ after $0.30$\,seconds, excluding $0.21$\,seconds of compilation,
with relative error $0.008$ in $\Sigma$. 
Following \cite{hong2026}, we also fit $\theta$ with Monte Carlo approximation of the 
likelihood by stochastic gradient descent with momentum for $1500$ steps.\footnote{In this likelihood evaluation, $\ptri$ is approximated as the mean over
$2000$ samples of $(u_1, u_2, v)$ in \eqref{eq:f_tri_int} of $1/(1 + e^{t/b})$, with
$t = \rho(u_1, u_2) - \min\{\rho(u_1, v), \rho(u_2, v)\}$ and $b = 5 \times 10^{-3}$, in place of the
indicator of $S_{\mathrm{tri}}$, so that $\ptri$ is differentiable in $\theta$.} 
The fit ends with $\log p(\htheta | y) - \log p(\theta | y) = 0.76$
after $2618$\,s, with relative error $0.029$.
With $\ptri$ in the log posterior evaluated instead by Monte Carlo with $10^5$ samples per trial, 
the two fits end $0.05$ and $0.85$ below $\htheta$  with standard errors of $0.07$ and $0.08$.

\section{Duo-trio probability}\label{sec:duotrio}
In this section, we introduce an algorithm for efficiently evaluating the Bernoulli probability
$\pduo$ of the duo-trio test (see \eqref{eq:f_duo_int}). 
We assume that $\Sx = \Sy$ for any two presented stimuli $x, y \in \mcX$.\footnote{This
assumption does not imply that the fitted covariance field is constant. Instead, 
we can view this assumption as an approximation whose error approaches zero when $x$ 
approaches $y$.} The case in which $\Sx \neq \Sy$ has already been handled in \cite{ennis1994}, 
for Euclidean $\rho$, which coincides with the form of \eqref{eq:dist}, after transforming 
$S$ to the identity matrix.
In what follows, when $\Sx = \Sy$, the duo-trio probability is reduced 
to an analytic function of one variable. The eigendecomposition and series summation of \cite{ennis1994} are
replaced by the evaluation of a Chebyshev expansion with coefficients computed 
once during precomputation, to a user-specified accuracy $\epsilon$.

In the Thurstonian model we consider for the duo-trio test, we have three 
Gaussians: $x_0, x_1$ i.i.d. distributed according to $\mcN(\mx, \Sigma)$, and 
independently, $y \sim \mcN(\my, \Sigma)$. 
The following is the central mathematical observation behind our algorithm.
We show that $\pduo$ depends on $\mx, \my, \Sigma$ entirely through the scalar
$r = (\my - \mx)^\top \Sigma^{-1} (\my - \mx)$. Furthermore, $\pduo$, as a function of 
$r$, is analytic and can be evaluated with a computationally tractable 
one-dimensional integral.
\begin{claim}\label{claim:duotrio_mahal}
For all $\mx, \my \in \R^{d}$ and all positive definite symmetric $d \times d$ matrices  $\Sigma$, 
\begin{align}\label{eq:duotrio_mahal}
\pduo(\mx, \my, \Sigma, \Sigma) =
\Pr\!\big({\chi'}^2_d(\lambda_1)>{\chi'}^2_d(\lambda_0)\big)
\end{align}
where
\begin{align}
   \lambda_1 = \tfrac{r}{6}\big(2 + \sqrt{3} \big), \quad \lambda_0 = \tfrac{r}{6} \big(2 - \sqrt{3} \big), 
  \quad r = (\my - \mx)^\top \Sigma^{-1} (\my - \mx)
\end{align}
where ${\chi'}^2_d(\lambda)$ denotes the noncentral
chi-squared distribution with $d$ degrees of freedom and noncentrality parameter $\lambda$.
\end{claim}
\begin{proof}
We first apply Claim \ref{claim:linear_invariance} to $\pduo(\mx, \my, \Sigma, \Sigma)$ 
with $A = \Sigma^{-1/2}$ and $b = -\Sigma^{-1/2}\mx$,
obtaining 
\begin{align}
\pduo(\mx, \my, \Sigma, \Sigma) = \pduo(0, \mu, I, I)
\end{align}
where
\begin{equation}
  \mu = \Sigma^{-1/2}(\my - \mx).
\end{equation}
We may therefore assume, without loss of generality, that 
$\mx = 0$, $\my = \mu$, $\Sigma = I$. 
We define $u_0, u_1$ to be the vectors 
\begin{equation}
  u_0 = x_1 - x_0, \qquad u_1 = y - x_0
\end{equation}
and define the multivariate Gaussian $\mbfu \in \R^{2d}$ by $\mbfu=[u_0\,\, u_1]$. Then 
\begin{equation}\label{eq:joint_uu}
  \mbfu \sim \mcN \!\left(\begin{bmatrix}0\\\mu\end{bmatrix},\, C \right),
  \qquad C = 
  \begin{bmatrix}2I & I\\ I & 2I\end{bmatrix}
\end{equation}
where $I$ is the $d\times d$ identity matrix.
Let $Q$ be the scalar 
\begin{align}
Q = \| u_0 \|^2 - \| u_1\|^2 
\end{align}
and observe that 
\begin{align}\label{eq:def_pduo_q}
\pduo(\mx, \my, \Sigma, \Sigma) = \Pr(Q < 0).
\end{align}
 Writing $Q$ as a quadratic form
\begin{equation}\label{eq:Q_form}
  Q = \mbfu^\top M \mbfu,
  \qquad
  M = \begin{bmatrix}I & \\ & -I\end{bmatrix}
\end{equation}
and using Lemma \ref{lem:duotrio}, we know
\begin{equation}\label{eq:Q_whitened}
  Q = \mbfz^\top C^{1/2} M C^{1/2} \mbfz
\end{equation}
where $\mbfz \sim \mcN(m, I)$ with $m$ defined in \eqref{eq:m_def}. 
Combining \eqref{eq:Q_form} and \eqref{eq:c12_def}, we observe that the matrix 
of the quadratic form \eqref{eq:Q_whitened} is diagonal. Specifically,
\begin{align}\label{eq:D_explicit}
  C^{1/2} M C^{1/2}
  &= \sqrt3 \begin{bmatrix}I & \\ & -I\end{bmatrix}.
\end{align}
Defining $\mbfz = [z_0 \,\, z_1]$ and substituting \eqref{eq:D_explicit} into \eqref{eq:Q_whitened}, we obtain
\begin{equation}\label{eq:Q_diag}
  Q = \mbfz^\top C^{1/2} M C^{1/2} \mbfz = \sqrt3\big(\|z_0\|^2 - \|z_1\|^2\big).
\end{equation}
Since $z_0, z_1$ are independent Gaussians with identity covariances, $\|z_0\|^2, \|z_1\|^2$ are
independent noncentral chi-squared distributions with $d$ degrees of freedom
whose noncentralities are (see \eqref{eq:m_def}),
\begin{equation}\label{eq:nc_explicit}
  \|\E [z_1] \|^2 = \|\mu\|^2 \Big(\tfrac{3+\sqrt3}{6}\Big)^{\!2}  = \| \mu \|^2 \frac{2+\sqrt3 }{6},
  \qquad
  \|\E [z_0] \|^2 = \|\mu\|^2 \Big(\tfrac{\sqrt3-3}{6}\Big)^{\!2}  = \| \mu \|^2 \frac{2-\sqrt3}{6}.
\end{equation}
Equation \eqref{eq:duotrio_mahal} follows immediately from combining \eqref{eq:def_pduo_q}, 
\eqref{eq:Q_diag}, \eqref{eq:nc_explicit}, and the fact that $\|\mu\|^2 = r$.
\end{proof}
Since the chi-squared distributions in equation \eqref{eq:duotrio_mahal} are independent, the 
p.d.f. of the joint distribution $\Big({\chi'}^2_d(\lambda_1), {\chi'}^2_d(\lambda_0)\Big)$
is the product of two one-dimensional p.d.f.s. Thus, 
\begin{equation}
\begin{aligned}\label{eq:pduo_int}
  \pduo(r)& = \int_0^\infty \int_t^\infty f_{d, \lambda_1(r)}(t') \,f_{d,\lambda_0(r)}(t) \, dt' \, dt
   = \int_0^\infty \overline F_{d,\lambda_1(r)}(t)\,f_{d,\lambda_0(r)}(t)\,dt
\end{aligned}
\end{equation}
where $f_{d,\lambda}$ denotes the p.d.f. of ${\chi'}^2_d(\lambda)$ (see, for example, \cite{johnson1995}), defined by 
\begin{equation}\label{eq:ncx2_pdf}
  f_{d,\lambda}(t) = \frac12 e^{-(t+\lambda)/2}
    \Big(\frac{t}{\lambda}\Big)^{\!\frac d4 - \frac12} I_{\frac d2 - 1}\!\big(\sqrt{\lambda t}\big),
\end{equation}
where $I_\nu$ is the modified Bessel function of the first kind
\cite[\href{https://dlmf.nist.gov/10.25.E2}{(10.25.2)}]{dlmf}, and where $\overline F_{d, \lambda}$
is defined by
\begin{equation}
  \overline F_{d,\lambda}(t) = \int_t^\infty f_{d,\lambda}(\tau)\,\mathrm{d}\tau,
\end{equation}
which is the so-called Marcum $Q$-function $Q_{d/2}(\sqrt\lambda, \sqrt t)$ \cite{marcum1950}.
Algorithms for the efficient evaluation of $f_{d, \lambda}$ and $\overline F_{d, \lambda}$
are implemented in standard software packages (see, for example, \cite{gil2014, benton2003}).
Thus, the problem of efficiently evaluating $\pduo(\mx, \my, \Sigma, \Sigma)$ for arbitrary arguments has been reduced to the evaluation of the one-dimensional integral \eqref{eq:pduo_int}.
Integral \eqref{eq:pduo_int} can be evaluated to floating point precision using adaptive 
quadrature rules (see, for example, \cite{trefethen2013}).

\subsection{Numerical evaluation}
Here we describe a numerical procedure for efficiently evaluating $\pduo$
with some error tolerance $\epsilon > 0$. 
We first observe that $\pduo(r) \to 1$ as $r \to \infty$ and 
choose a truncation $R$ such that $|\pduo(r) - 1| < \epsilon / 10$ for all $r > R$.
We then construct a Chebyshev expansion of order $n$ on the interval $[0, R]$ 
such that coefficients have decayed to a magnitude smaller than $\epsilon / 10$
by the $n$th coefficient. 
Since $\pduo$ is analytic, decay of the coefficients in 
a Chebyshev expansion is exponential \cite{trefethen2013} and thus achieving 
$\epsilon$ accuracy requires $\mathcal{O}(-\log \epsilon)$ terms for fixed $R$.
Specifically, we choose the order of the Chebyshev expansion, $n$, 
then evaluate $\pduo(x_i)$ for $i=1,\dots,n$ where $x_1,\dots,x_n$ are the 
order-$n$ Chebyshev nodes \cite{trefethen2013} appropriately transformed 
to the interval $[0, R]$. We evaluate $\pduo(x_i)$ with adaptive Gauss--Kronrod 
quadrature \cite{piessens1983,virtanen2020}, at tolerance $\epsilon / 10$. 

We then compute the corresponding coefficients in the Chebyshev expansion. 
If the coefficients have not yet decayed to below $\epsilon / 10$ in magnitude, we 
use larger $n$ and repeat this procedure. 
We store the $n$ coefficients in the Chebyshev expansion. We can then evaluate 
$\pduo$ for arbitrary inputs, to $\epsilon$ accuracy by evaluating the same 
$n$-degree Chebyshev expansion at arbitrary $r \in [0, R]$. 
This procedure produces a fast and accurate evaluation of $\pduo$.  
The evaluation is amenable to automatic differentiation and just-in-time compilation \cite{jax2018}.

\begin{figure}[h]
\centering
\includegraphics[width=0.49\textwidth]{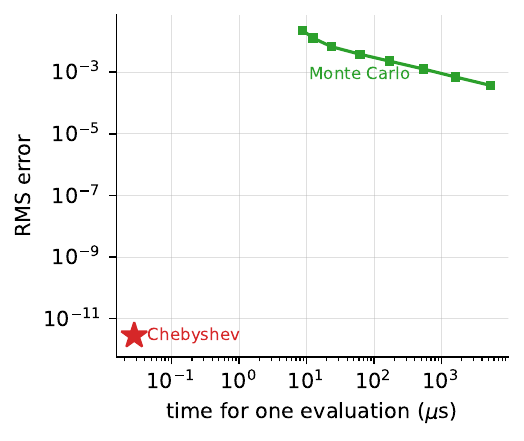}
\caption{Root mean square error of evaluation of $\pduo$ for $d = 3$. 
We compare the Chebyshev expansion with $\epsilon = 10^{-8}$ (red) and 
Monte Carlo approximation for $n$ from $30$ to $10^{5}$ (green).}
\label{fig:duotrio_plot}
\end{figure}

In Figure \ref{fig:duotrio_plot} we compare evaluation of $\pduo$ using our numerical
procedure for $d=3$, and $\epsilon = 10^{-8}$, to Monte Carlo with 
various numbers of samples. 
We report root mean square error over $400$ points in $[0, R]$, where ground truth 
is computed to machine precision using adaptive quadrature. 
The Chebyshev expansion takes $0.028\,\mu$s per evaluation,
with sup norm error $8.7 \times 10^{-12}$ on $[0, R]$, measured in a self-convergence 
sense. That is, we use as reference a Chebyshev expansion with a large number of terms
and with coefficients evaluated to near floating point precision.
Adaptive quadrature at tolerance
$10^{-10}$ took $28$\,ms per evaluation using $15$-point Gauss--Kronrod quadrature
 \cite{piessens1983}.
The approach of \cite[eq.~(11)]{ennis1994} also avoids quadrature for each evaluation of 
$\pduo$. However, the number of terms needed for error $\epsilon$ grows with $r$, since the 
noncentralities \eqref{eq:nc_explicit} are proportional to $r$. At $r = R$ \cite{ennis1994} 
requires $2328$ terms for error at most $10^{-8}$, compared to $36$ for the algorithm of this 
section.
At $\epsilon = 10^{-8}$, our procedure gives $R = 220.8$ and $n = 36$ 
terms for $d = 3$. For $d = 2$ we obtain $R = 218.4$ 
and $n = 36$. 
The timings in Figure \ref{fig:duotrio_plot} do not include the precomputation step 
which is done once for any given $d$ and $\epsilon$.

\section{Discussion}\label{sec:discussion}
We introduced algorithms for evaluating the Bernoulli probabilities $\ptri$ and $\pduo$ 
of \eqref{eq:p_triangle}, \eqref{eq:p_duotrio} that are substantially more efficient than 
existing methods. 
Our methods enable efficient statistical inference and other computational tasks related
to Thurstonian perception models. For example, when using our approximation 
of $\ptri$ on the Bayesian model of \cite{hong2026}, we obtain a more accurate MAP
estimate than with Monte Carlo evaluation in over three orders 
of magnitude less time. 
Because the approximations are efficient, deterministic, and differentiable, they also 
make it practical to sample the posterior with gradient-based samplers such as the no-U-turn 
sampler (NUTS) \cite{hoffman2014}, and to place trials adaptively, which requires posterior
inference in the time between trials.

For both $\ptri, \pduo$, the central mathematical observations underlying the efficient 
evaluation are (i) an analytic transformation from a function of $\mathcal{O}(d^2)$ 
arguments to $\mathcal{O}(d)$ arguments, and (ii) the lower dimensional functions 
are smooth and can be efficiently approximated with 
neural networks ($\ptri$) or special function expansions ($\pduo$). 
For $\ptri$, the gain comes from both the transformation and the network --- on the 
original arguments, a neural network needs two orders of magnitude more parameters 
and roughly seven times more time for the same accuracy (Figure \ref{fig:reduction}).

The reduction of variables we use relies on two properties of the model, (i) the noise 
family (here Gaussian) is closed under affine transformation of the stimulus space, 
and (ii) the response depends on the internal representations only 
through comparisons of $\rho$ of the form \eqref{eq:dist}. It therefore applies to noise 
with elliptical level sets other than Gaussian, and to tests with more stimuli per trial, such as the 
tetrad test \cite{ennis1998}, in which four stimuli, two of each kind, are sorted into two 
pairs. 
The same reduction of variables can be used for the Bernoulli probability of the 
same--different test \cite{brockhoff2010, sensR}. In the same--different test, a
subject is presented with two stimuli and reports whether they are the same.  
Internal representations are $x \sim \mcN(\mx, \Sx)$ and $y \sim \mcN(\my, \Sy)$, 
and the subject is modeled as reporting that the stimuli differ when 
$\rho(x, y) > \tau$, for a fixed $\tau > 0$.

The network approximation of $h$ has several limitations. 
Its error is estimated on a held out set rather than bounded analytically. 
The accuracy of the network depends on the optimization procedure, on the 
number and distribution of training points, and on the accuracy of the training labels. 
Additionally, a separate network (or special function expansion) needs to be 
computed for each dimensionality, $d$, and error tolerance $\epsilon$.

\section*{Acknowledgements}
We thank Manas Rachh for many helpful discussions about algorithm development. 
Portions of the software development, numerical experiments, and editing 
of this paper were carried out with the assistance of AI agents.

\bibliographystyle{plain}
\bibliography{refs_overleaf}

\begin{thebibliography}{10}

\bibitem{amazeen1999perceptual}
Eric~L. Amazeen.
\newblock Perceptual independence of size and weight by dynamic touch.
\newblock {\em Journal of Experimental Psychology: Human Perception and
  Performance}, 25(1):102--119, 1999.

\bibitem{ashby1986varieties}
F.~Gregory Ashby and James~T. Townsend.
\newblock Varieties of perceptual independence.
\newblock {\em Psychological Review}, 93(2):154--179, 1986.

\bibitem{benton2003}
Denise Benton and K.~Krishnamoorthy.
\newblock Computing discrete mixtures of continuous distributions: noncentral
  chisquare, noncentral $t$ and the distribution of the square of the sample
  multiple correlation coefficient.
\newblock {\em Computational Statistics \& Data Analysis}, 43(2):249--267,
  2003.

\bibitem{jax2018}
James Bradbury, Roy Frostig, Peter Hawkins, Matthew~James Johnson, Yash
  Katariya, Chris Leary, Dougal Maclaurin, George Necula, Adam Paszke, Jake
  Vander{P}las, Skye Wanderman-{M}ilne, and Qiao Zhang.
\newblock {JAX}: composable transformations of {P}ython+{N}um{P}y programs,
  2018.
\newblock \url{http://github.com/jax-ml/jax}.

\bibitem{brockhoff2010}
Per~B. Brockhoff and Rune Haubo~Bojesen Christensen.
\newblock Thurstonian models for sensory discrimination tests as generalized
  linear models.
\newblock {\em Food Quality and Preference}, 21(3):330--338, 2010.

\bibitem{sensR}
Rune Haubo~Bojesen Christensen and Per~B. Brockhoff.
\newblock {\em {sensR}---An {R}-package for sensory discrimination}, 2023.
\newblock R package version 1.5-3.

\bibitem{dlmf}
{\it NIST Digital Library of Mathematical Functions}.
\newblock \url{https://dlmf.nist.gov/}, Release 1.2.8 of 2026-09-15.
\newblock F.~W.~J. Olver, A.~B. {Olde Daalhuis}, D.~W. Lozier, B.~I. Schneider,
  R.~F. Boisvert, C.~W. Clark, B.~R. Miller, B.~V. Saunders, H.~S. Cohl, and
  M.~A. McClain, eds.

\bibitem{docarmo1992}
Manfredo~P. do~Carmo.
\newblock {\em Riemannian Geometry}.
\newblock Birkh\"auser, Boston, 1992.

\bibitem{ennis1993}
Daniel~M. Ennis.
\newblock The power of sensory discrimination methods.
\newblock {\em Journal of Sensory Studies}, 8(4):353--370, 1993.

\bibitem{ennis1994}
Daniel~M. Ennis and Norman~L. Johnson.
\newblock A general model for preferential and triadic choice in terms of
  central {F} distribution functions.
\newblock {\em Psychometrika}, 59(1):91--96, 1994.

\bibitem{ennis1985}
Daniel~M. Ennis and Kenneth Mullen.
\newblock The effect of dimensionality on results from the triangular method.
\newblock {\em Chemical Senses}, 10(4):605--608, 1985.

\bibitem{ennis1986}
Daniel~M. Ennis and Kenneth Mullen.
\newblock A multivariate model for discrimination methods.
\newblock {\em Journal of Mathematical Psychology}, 30:206--219, 1986.

\bibitem{ennis1998}
John~M. Ennis, Daniel~M. Ennis, Dorene Yip, and Michael O'Mahony.
\newblock Thurstonian models for variants of the method of tetrads.
\newblock {\em British Journal of Mathematical and Statistical Psychology},
  51(2):205--215, 1998.

\bibitem{fechner1860}
Gustav~Theodor Fechner.
\newblock {\em Elemente der Psychophysik}.
\newblock Breitkopf und H{\"a}rtel, Leipzig, 1860.

\bibitem{fisher1935}
Ronald~A. Fisher.
\newblock {\em The Design of Experiments}.
\newblock Oliver and Boyd, Edinburgh, 1935.

\bibitem{fitousi2013variants}
Daniel Fitousi and Michael~J. Wenger.
\newblock Variants of independence in the perception of facial identity and
  expression.
\newblock {\em Journal of Experimental Psychology: Human Perception and
  Performance}, 39(1):133--155, 2013.

\bibitem{geisler2011}
Wilson~S. Geisler.
\newblock Contributions of ideal observer theory to vision research.
\newblock {\em Vision Research}, 51(7):771--781, 2011.

\bibitem{gil2014}
Amparo Gil, Javier Segura, and Nico~M. Temme.
\newblock Algorithm 939: Computation of the {Marcum} {Q}-function.
\newblock {\em ACM Transactions on Mathematical Software}, 40(3):20:1--20:21,
  2014.

\bibitem{green1966}
David~M. Green and John~A. Swets.
\newblock {\em Signal Detection Theory and Psychophysics}.
\newblock Wiley, New York, 1966.

\bibitem{hartley2022}
Isabella Hartley, Liliana Orellana, Djin~Gie Liem, and Russell Keast.
\newblock Assessment of the triangle test methodology for determining umami
  discrimination status.
\newblock {\em Chemical Senses}, 47:bjac003, 2022.

\bibitem{hoffman2014}
Matthew~D. Hoffman and Andrew Gelman.
\newblock The no-{U}-turn sampler: Adaptively setting path lengths in
  {H}amiltonian {M}onte {C}arlo.
\newblock {\em Journal of Machine Learning Research}, 15:1593--1623, 2014.

\bibitem{hong2026}
Fangfang Hong, Ruby Bouhassira, Jason Chow, Craig Sanders, Michael Shvartsman,
  Phillip Guan, Alex~H Williams, and David~H Brainard.
\newblock Comprehensive characterization of human color discrimination
  thresholds.
\newblock {\em eLife}, 14:RP108943, July 2026.

\bibitem{hummel2007}
Thomas Hummel, G.~Kobal, H.~Gudziol, and A.~Mackay-Sim.
\newblock Normative data for the ``{Sniffin' Sticks}'' including tests of odor
  identification, odor discrimination, and olfactory thresholds: an upgrade
  based on a group of more than 3,000 subjects.
\newblock {\em European Archives of Oto-Rhino-Laryngology}, 264(3):237--243,
  2007.

\bibitem{hummel1997}
Thomas Hummel, B.~Sekinger, S.~R. Wolf, E.~Pauli, and G.~Kobal.
\newblock `{Sniffin' Sticks}': olfactory performance assessed by the combined
  testing of odor identification, odor discrimination and olfactory threshold.
\newblock {\em Chemical Senses}, 22(1):39--52, 1997.

\bibitem{iso4120}
International Organization for Standardization.
\newblock {\em Sensory analysis --- Methodology --- Triangle test}, 2021.
\newblock ISO 4120:2021.

\bibitem{iso10399}
International Organization for Standardization.
\newblock {\em Sensory analysis --- Methodology --- Duo-trio test}, 2026.
\newblock ISO 10399:2026.

\bibitem{johnson1995}
Norman~L. Johnson, Samuel Kotz, and N.~Balakrishnan.
\newblock {\em Continuous Univariate Distributions}, volume~2.
\newblock John Wiley \& Sons, New York, 2 edition, 1995.

\bibitem{kapenga1987}
John~A. Kapenga, Kenneth Mullen, Elise de~Doncker, and Daniel~M. Ennis.
\newblock The integration of the multivariate normal density function for the
  triangular method.
\newblock In Patrick Keast and Graeme Fairweather, editors, {\em Numerical
  Integration: Recent Developments, Software and Applications}, volume 203 of
  {\em NATO ASI Series C: Mathematical and Physical Sciences}, pages 321--328.
  D. Reidel, Dordrecht, 1987.

\bibitem{kim2010}
Min-A Kim, Young-Mi Lee, and Hye-Seong Lee.
\newblock Comparison of $d'$ estimates produced by three versions of a duo-trio
  test for discriminating tomato juices with varying salt concentrations: the
  effects of the number and position of the reference stimulus.
\newblock {\em Food Quality and Preference}, 21(5):504--511, 2010.

\bibitem{kingma2015}
Diederik~P. Kingma and Jimmy Ba.
\newblock Adam: A method for stochastic optimization.
\newblock In {\em 3rd International Conference on Learning Representations,
  {ICLR} 2015}, 2015.

\bibitem{macadam1942}
David~L. MacAdam.
\newblock Visual sensitivities to color differences in daylight.
\newblock {\em Journal of the Optical Society of America}, 32(5):247--274,
  1942.

\bibitem{macmillan2005detection}
Neil~A. Macmillan and C.~Douglas Creelman.
\newblock {\em Detection Theory: A User's Guide}.
\newblock Lawrence Erlbaum Associates, Mahwah, NJ, 2 edition, 2005.

\bibitem{marcum1950}
Jess~I. Marcum.
\newblock Table of {Q} functions.
\newblock Research Memorandum RM-339, The RAND Corporation, Santa Monica, CA,
  1950.

\bibitem{mcmahon2016}
Emma McMahon, Rozlynne Clarke, Rachael Jaenke, and Julie Brimblecombe.
\newblock Detection of 12.5\% and 25\% salt reduction in bread in a remote
  {I}ndigenous {A}ustralian community.
\newblock {\em Nutrients}, 8(3):169, 2016.

\bibitem{morimoto2026}
Takuma Morimoto.
\newblock Seeing colour and material in real-world lighting environments.
\newblock Talk presented at the International Colour Vision Symposium,
  Brighton, UK, 2026.
\newblock August 14--18, 2026.

\bibitem{mullen1987}
Kenneth Mullen and Daniel~M. Ennis.
\newblock Mathematical formulation of multivariate {E}uclidean models for
  discrimination methods.
\newblock {\em Psychometrika}, 52(2):235--249, 1987.

\bibitem{mullen1991}
Kenneth Mullen and Daniel~M. Ennis.
\newblock A simple multivariate probabilistic model for preferential and
  triadic choices.
\newblock {\em Psychometrika}, 56(1):69--75, 1991.

\bibitem{mullen1988}
Kenneth Mullen, Daniel~M. Ennis, Elise de~Doncker, and John~A. Kapenga.
\newblock Models for the duo-trio and triangular methods.
\newblock {\em Biometrics}, 44(4):1169--1175, 1988.

\bibitem{piessens1983}
Robert Piessens, Elise de~Doncker-Kapenga, Christoph~W. {\"U}berhuber, and
  David~K. Kahaner.
\newblock {\em {QUADPACK}: A Subroutine Package for Automatic Integration}.
\newblock Springer-Verlag, Berlin, 1983.

\bibitem{saavedra2015}
Lorena Saavedra-Garcia, Antonio Bernabe-Ortiz, Robert~H. Gilman, Francisco
  Diez-Canseco, Mar{\'i}a~Kathia C{\'a}rdenas, Katherine~A. Sacksteder, and
  J.~Jaime Miranda.
\newblock Applying the triangle taste test to assess differences between low
  sodium salts and common salt: evidence from {P}eru.
\newblock {\em PLOS ONE}, 10(7):e0134700, 2015.

\bibitem{super2026}
{Schematic}.
\newblock Super, 2026.
\newblock Python package \texttt{schematic-supertest}, version 0.1.1,
  \url{https://docs.schematic.tech/super}, accessed 2026-11-02.

\bibitem{soto2015general}
Fabian~A. Soto, Lauren Vucovich, Robert Musgrave, and F.~Gregory Ashby.
\newblock General recognition theory with individual differences: A new method
  for examining perceptual and decisional interactions with an application to
  face perception.
\newblock {\em Psychonomic Bulletin \& Review}, 22(1):88--111, 2015.

\bibitem{thurstone1927}
L.~L. Thurstone.
\newblock A law of comparative judgment.
\newblock {\em Psychological Review}, 34(4):273--286, 1927.

\bibitem{trefethen2013}
Lloyd~N. Trefethen.
\newblock {\em Approximation Theory and Approximation Practice}.
\newblock SIAM, Philadelphia, 2013.

\bibitem{virtanen2020}
Pauli Virtanen, Ralf Gommers, Travis~E. Oliphant, et~al.
\newblock {SciPy} 1.0: fundamental algorithms for scientific computing in
  {Python}.
\newblock {\em Nature Methods}, 17(3):261--272, 2020.

\end{thebibliography}

\appendix
\section{The ideal observer $\rho$}\label{sec:ml_rule}
The model of \cite{hong2026} for the triangle test compares the internal representations
$x_1, x_2 \sim \mcN(\mx, \Sx)$ and $y \sim \mcN(\my, \Sy)$ through their three pairwise distances
under the distance function
\begin{align}\label{eq:rho_appendix}
\rho(x, y) = (x - y)^\top S^{-1} (x - y)
\end{align}
where $S = \tfrac{2}{3} \Sx + \tfrac{1}{3} \Sy$. 
Thus their decision rule assumes knowledge of the internal representations $x_1, x_2, y$
and $\Sx, \Sy$, but not the true stimuli themselves $\mx, \my$.  
Here, under those same assumptions, we derive the decision rule 
for the ideal observer \cite{green1966, geisler2011} for both the triangle and duo-trio 
tasks. We show that such an ideal observer uses the same decision rule, with $S$ of 
\eqref{eq:rho_appendix} replaced with $S = \Sx$.
While a real observer may not have access to $\Sx, \Sy$, after having been presented
with enough stimuli in the neighborhood of $\mx, \my$, the observer may have a 
reasonable guess of $\Sx, \Sy$. 
Ideal observers have proven useful in vision research, both as a benchmark for human 
performance and as a principled model of real observers~\cite{geisler2011}.

We start by assuming that an ideal observer who knows $\Sx, \Sy$
observes the internal representations $w_0, w_1, w_2$,
where $w_i, w_j \sim \mcN(\mx, \Sx), w_k \sim \mcN(\my, \Sy)$ for some $i, j, k$
such that $\{i, j, k\} = \{0, 1, 2\}$. We assume that the ideal observer does 
not know which is the unique stimulus. That is, the ideal observer does not know 
$k \in \{0, 1, 2\}$ and $\{i, j\} = \{0, 1, 2\} \setminus \{k\}$. 

Under our model, which we assume the ideal observer knows, we have three
unknown parameters --- $\mx, \my$, and the index of the unique stimulus $k \in \{0, 1, 2\}$.
For any $k \in \{0, 1, 2\}$ and $\{i, j\} = \{0, 1, 2\} \setminus \{k\}$, the likelihood
that $w_k$ is the internal representation of the unique stimulus is given by
\begin{align}\label{eq:lik_kappa}
p(w_0, w_1, w_2 \,|\, k, \mx, \my) = \phi(w_i; \mx, \Sx)\, \phi(w_j; \mx, \Sx)\, \phi(w_k; \my, \Sy).
\end{align}
The ideal observer with uniform prior on $k \in \{0, 1, 2\}$ and uniform (improper) 
prior on $\mx, \my$ chooses
\begin{align}\label{eq:khat_def}
\ktri := \arg\max_{\kappa} \int_{\R^d} \int_{\R^d} p(w_0, w_1, w_2 \,|\, \kappa, \mx, \my) \, d\my \, d\mx.
\end{align}
While the uniform prior on $\R^d$ is improper, the integral \eqref{eq:khat_def} has a Gaussian integrand
and is thus finite. With the information known by the ideal observer, the choice of
stimulus $\ktri$ maximizes the posterior probability of identifying the unique stimulus.

In the duo-trio test, we have a reference stimulus, whose internal representation we denote 
$w_0 \sim \mcN(\mx, \Sx)$, and two further stimuli with internal representations $w_1, w_2$
where $w_{k} \sim \mcN(\mx, \Sx)$, $w_{\notk} \sim \mcN(\my, \Sy)$ for 
$\{ k, \notk \} = \{1, 2\}$. The likelihood that $w_k$ is the internal 
representation of the repeated copy of the reference is given by
\begin{align}
p(w_1, w_2 \,|\, w_0, k, \mx, \my) = 
\phi(w_0; \mx, \Sx)\, \phi(w_k; \mx, \Sx)\, \phi(w_{\notk}; \my, \Sy),
\end{align}
and the ideal observer
with uniform prior on $k, \mx, \my$ chooses
\begin{align}\label{eq:khat_duo_def}
\hat k^{\mathrm{duo}} := \arg\max_{\kappa \in \{1, 2\}} \int_{\R^d} \int_{\R^d}
\phi(w_0; \mx, \Sx)\, \phi(w_\kappa; \mx, \Sx)\, \phi(w_{\notkappa}; \my, \Sy) \, d\my \, d\mx.
\end{align}

\begin{definition}
The ideal observer's response is $\ktri$ of \eqref{eq:khat_def} for the triangle test and 
$\hat k^{\mathrm{duo}}$ of \eqref{eq:khat_duo_def} for the duo-trio test.
\end{definition}

The following, which is the main result of this section, shows that $\ktri$ and
$\hat k^{\mathrm{duo}}$ are the choices made by the rules of \eqref{eq:p_triangle} and
\eqref{eq:p_duotrio} with $S = \Sx$, and hence that $\ptri$ and $\pduo$ are the ideal
observer's probabilities of a correct response.
\begin{claim}\label{claim:ml_rule}
For all $w_0, w_1, w_2 \in \R^d$, the unique stimulus selected by the ideal observer, $\ktri$ (see \eqref{eq:khat_def}), for the triangle test satisfies
\begin{align}\label{eq:ml_argmax}
\ktri = \operatorname*{argmin}_{\kappa \in \{0, 1, 2\}} (w_i - w_j)^\top \Sx^{-1} (w_i - w_j)
\end{align}
where $\kappa \in  \{0, 1, 2\}$ and $\{i, j\} = \{0, 1, 2\} \setminus \{\kappa\}$.
The choice $\hat k^{\mathrm{duo}}$ (see \eqref{eq:khat_duo_def}) for the duo-trio test satisfies
\begin{align}\label{eq:ml_argmax_duo}
\hat k^{\mathrm{duo}} = \operatorname*{argmin}_{\kappa \in \{1, 2\}} (w_0 - w_\kappa)^\top \Sx^{-1} (w_0 - w_\kappa).
\end{align}
Moreover, for $\rho$ of the form \eqref{eq:dist} with $S = \Sx$,
\begin{align}\label{eq:ml_prob}
\ptri(\mx, \my, \Sx, \Sy) &= \Pr(\ktri = k), \\
\pduo(\mx, \my, \Sx, \Sy) &= \Pr(\hat k^{\mathrm{duo}} = k).
\end{align}
\end{claim}
\begin{proof}
Fix $w_0, w_1, w_2 \in \R^d$ and $\kappa \in \{0, 1, 2\}$, and let $\{i, j\} = \{0, 1, 2\} \setminus \{\kappa\}$.
It follows immediately from the definition of the Gaussian probability density function $\phi$
that 
\begin{align}\label{eq:inner_int}
\int_{\R^d} \phi(w_\kappa; \my, \Sy)\, d\my = \int_{\R^d} \phi(\my; w_\kappa, \Sy)\, d\my = 1.
\end{align}
Substituting \eqref{eq:inner_int} and \eqref{eq:lik_kappa} into \eqref{eq:khat_def} we have 
\begin{align}
\ktri = \arg\max_{\kappa} \int_{\R^d} \phi(w_i; \mx, \Sx)\, \phi(w_j; \mx, \Sx)\, \, d\mx.
\end{align}
Expanding the Gaussian p.d.f. $\phi$, and observing that the normalizing constant is 
independent of $\kappa$, we obtain
\begin{align}
\ktri = \arg\max_{\kappa} \int_{\R^d} \exp\bigg( -\tfrac{1}{2} \bigl[ (w_i - \mx)^\top \Sx^{-1} (w_i - \mx) + (w_j - \mx)^{\top} \Sx^{-1} (w_j - \mx) \bigr] \bigg) \, d\mx
\end{align}
We split the Gaussian integrand into the product of 
one Gaussian in $\mx$ and another independent of $\mx$, 
\begin{align}
\ktri & = \arg\max_{\kappa} \int_{\R^d} \exp\Bigl( -\tfrac12 \bigl[ (w_i - \mx)^\top \Sx^{-1} (w_i - \mx)
  + (w_j - \mx)^\top \Sx^{-1} (w_j - \mx) \bigr] \Bigr) d\mx \\
& = \arg\max_{\kappa} \int_{\R^d} \exp\Bigl( -\tfrac14 (w_i - w_j)^\top \Sx^{-1} (w_i - w_j)
  \; - (\mx - \alpha)^\top \Sx^{-1} (\mx - \alpha) \Bigr) d\mx \\
& = \arg\max_{\kappa} \,\, \exp\Bigl( -\tfrac14 (w_i - w_j)^\top \Sx^{-1} (w_i - w_j) \Bigr)
\, \int_{\R^d}   \exp\Bigl( - (\mx - \alpha)^\top \Sx^{-1} (\mx - \alpha) \Bigr) d\mx  \\
& = \arg\max_{\kappa} \,\, \exp\Bigl( -\tfrac14 (w_i - w_j)^\top \Sx^{-1} (w_i - w_j) \Bigr)
\end{align}
where $\alpha = (w_i + w_j)/2$.

For the duo-trio test, the same analysis with $(w_i, w_j, w_\kappa)$ replaced by
$(w_0, w_\kappa, w_{\notkappa})$ shows that the integral in \eqref{eq:khat_duo_def} equals
$\det(4\pi\Sx)^{-1/2} \exp\bigl(-\tfrac14 (w_0 - w_\kappa)^\top \Sx^{-1} (w_0 - w_\kappa)\bigr)$,
which gives \eqref{eq:ml_argmax_duo}.

For \eqref{eq:ml_prob}, set $(w_i, w_j, w_k) = (x_1, x_2, y)$. By \eqref{eq:ml_argmax},
$\ktri = k$ if and only if $\rho(x_1, x_2) < \min\{\rho(x_1, y), \rho(x_2, y)\}$, the same event
as \eqref{eq:p_triangle}. Likewise, with
$(w_0, w_k, w_{\notk}) = (x_{\mathrm{ref}}, x_1, y_1)$, \eqref{eq:ml_argmax_duo} gives
$\hat k^{\mathrm{duo}} = k$ if and only if $\rho(x_{\mathrm{ref}}, x_1) < \rho(x_{\mathrm{ref}}, y_1)$,
the event in \eqref{eq:p_duotrio}.
\end{proof}

\begin{remark}
The choice of $\ktri$ in \eqref{eq:ml_argmax} remains unchanged when 
the integrals in \eqref{eq:khat_def} are replaced by maxima over $\mx, \my$. The same holds for the duo-trio test.
Indeed, for every $\kappa$, 
\begin{align*}
\max_{\mx, \my} &p(w_0, w_1, w_2 \,|\, \kappa, \mx, \my) \\
& = \det(\pi\Sx)^{-1/2}\det(2\pi\Sy)^{-1/2} \int_{\R^d} \int_{\R^d} p(w_0, w_1, w_2 \,|\, \kappa, \mx, \my) \, d\my \, d\mx,
\end{align*}
where the constant does not depend on $\kappa$.
\end{remark}

\end{document}